\documentclass[useAMS,usenatbib]{mn2e}

\usepackage{amsmath, amssymb}
\usepackage[dvips]{graphicx}
\usepackage{epsfig}
\usepackage{color}

\newcommand{\rot}{\bmath{\nabla} \times}
\newcommand{\divg}{\bmath{\nabla}\cdot}

\newcommand{\rlight}{r_{\rm L}}

\newcommand{\BQ}{B_{\rm Q}}

\newcommand{\me}{m_{\rm e}}

\newcommand{\aap}{A\&A}
\newcommand{\mnras}{MNRAS}
\newcommand{\apss}{Ap\&SS}

\newcommand{\apj}{ApJ}
\newcommand{\apjl}{ApJL}
\newcommand{\apjs}{ApJS}

\newcommand{\prd}{Physical Review D}

\newcommand{\prb}{Phys. Rev. B}
\newcommand{\physrep}{Phys. Rep.}
\newcommand{\jcap}{Jour. Cosm. Astro. Phys.}

\title[GRFFQED fields]{A 3+1 formalism for quantum electrodynamical corrections to Maxwell equations in general relativity}
\author[J. P\'etri]{J.  P\'etri$^{1}$
\thanks{E-mail: jerome.petri@astro.unistra.fr} \\
  $^{1}$Observatoire astronomique de Strasbourg, Universit\'e de Strasbourg, CNRS, UMR 7550, 11 rue de l'universit\'e, F-67000 Strasbourg, France.}
\begin{document}

\date{Accepted . Received ; in original form }

\pagerange{\pageref{firstpage}--\pageref{lastpage}} 

\maketitle

\label{firstpage}

\begin{abstract}
Magnetized neutron stars constitute a special class of compact objects harbouring gravitational fields that deviate strongly from the Newtonian weak field limit. Moreover strong electromagnetic fields anchored into the star give rise to non-linear corrections to Maxwell equations described by quantum electrodynamics (QED). Electromagnetic fields close to or above the critical value of $\BQ=4.4\times10^9$~T are probably present in some pulsars and for most of the magnetars. To account properly for emission emanating from the neutron star surface like for instance thermal radiation and its polarization properties, it is important to include general relativistic (GR) effects simultaneously with non-linear electrodynamics. This can be achieved through a 3+1 formalism known in general relativity and that incorporates QED perturbations to Maxwell equations. Starting from the lowest order corrections to the Lagrangian for the electromagnetic field, as given for instance by Born-Infeld or Euler-Heisenberg theory, we derive the non-linear Maxwell equations in general relativity including quantum vacuum effects. We also derive a prescription for the force-free limit and show that these equations can be solved with classical finite volume methods for hyperbolic conservation laws. It is therefore straightforward to include general relativity and quantum electrodynamics in the description of neutron star magnetospheres by using standard classical numerical techniques borrowed from Maxwell and Newton theory. As an application, we show that spin-down luminosity corrections associated to QED effects are negligible with respect to GR corrections.
\end{abstract}


\section{Introduction}
\label{sec:Introduction}

As a final stage of the stellar evolution process, neutron stars 
constitute a special class of compact objects showing strong field effects in 
both gravitational and electromagnetic interactions. Typical values for 
their radius lye around $R\approx12$~km whereas their fiducial mass is 
about $M=1.4$~$M_\odot$ (where $M_\odot$ is the mass of the sun) thus 
leading to a compactness parameter  
defined by \begin{equation}
 \Xi = 2\,G\,M/R\,c^2 \approx 0.34 \;.
\end{equation}
This compactness is close to the maximal reachable compactness
defining a black hole. $G$ is the gravitational constant and $c$ the speed of 
light. General relativity is therefore required for an accurate description of 
the gravitational field in the vicinity of the neutron star. Moreover, from 
simple estimates of their spin-down luminosity being interpreted as 
magneto-dipole losses, normal pulsars can harbour magnetic fields of strength 
comparable to $B\approx10^8$~T and even higher fields for magnetars, of the 
order $B\approx10^{10-11}$~T. About thirty magnetars are known today and 
compiled in a catalogue described in \cite{2014ApJS..212....6O}. These values 
are comparable or largely above the critical field value of 
\begin{equation}
 \BQ = \frac{\me\,c^2}{e\,\hbar} \approx 4.4\times10^9 \textrm{ T} .
\end{equation}
$\me$ is the mass of an electron, $e$ the absolute value of its electric 
charge and $\hbar$ the reduced Planck constant.
At such field strengths, quantum electrodynamical corrections to  
classical electromagnetism as given by Maxwell equations become significant. 
Vacuum polarization and electron-positron pair creation are two examples of such 
QED effects. The perturbations induced by QED can be conveniently 
expressed in terms of an effective Lagrangian field theory such as the one found 
empirically by \cite{1934RSPSA.144..425B} or derived directly from a first order 
expansion of the QED Lagrangian as given by \cite{1936ZPhy...98..714H}. Such 
effective Lagrangians are very fruitful in getting more physical insight and 
intuition on QED effects and represent very efficient tools to extend classical 
electrodynamics methods to the realm of quantum physics. Extreme fields 
around $\BQ$ cannot be reach in terrestrial laboratories although the power of 
current lasers are approaching it in order to do reproducible experiments and 
check the theory \citep{2014EPJST.223.1063K}. Pulsars and magnetars could be 
used as extraterrestrial laboratories to investigate QED in strong 
electromagnetic fields and in curved space-time. The electromagnetic 
properties of quantum vacuum in special relativity are summarized in 
\cite{2013RPPh...76a6401B}.
\cite{2010PhR...487....1R} propose a comprehensive review on electron-positron 
pair creation/annihilation in the physical and astrophysical contexts. The 
classical picture of high-energy radiation processes in astrophysics need also 
to be revised according to QED corrections. This is discusses in 
\cite{2006RPPh...69.2631H} with a special emphasize to neutron star interiors 
and atmospheres. For instance, \cite{2005MNRAS.362..777H, 2007Ap&SS.308..101H} 
suggested an explanation of the non-thermal emission from magnetar in terms of 
QED.

Not only radiation is affected by QED but also the behaviour of 
pair plasmas and electron-ion plasmas as explained in depth in the review by 
\cite{2014RPPh...77c6902U}. The plasma dielectric tensor and therefore also the 
normal modes of propagation of electromagnetic waves and plasma oscillations are 
affected. This had led some authors to extend the equations of fluid 
hydrodynamics to some quantum aspects. For instance 
\cite{2001PhRvB..64g5316M} looked at some of these corrections and 
\cite{2005PhPl...12f2117H} added the magnetic field into the description
leading to quantum plasma magnetohydrodynamic (QMHD). QED corrections to MHD 
have already been proposed by \cite{1998PhRvD..57.3219T} and by 
\cite{1999PhRvD..59d5005H}. Interestingly, the non-linear nature of 
electrodynamics in strong fields induces shocks as in the case of flows in 
hydrodynamics as pointed out by \cite{1998PhRvD..58d3005H}, see also 
\cite{2011MNRAS.412.1381M} for non-linear wave propagation in the magnetospheric 
plasma. From a more fundamental point of view, QED corrections can be brought 
to any magnetic multipole, like for instance the QED corrections to the
magnetic dipole as investigated by \cite{1997JPhA...30.6475H}.

Gravitational effects are usually ignored. Nevertheless 
\cite{2001PhRvD..63f4028H} applied quantum field theory near a rotating black 
hole and showed that it could copiously produce electron-positron pairs. Also 
\cite{2013PhRvD..88h5004R} studied QED corrections in strong gravity by 
inspecting spherically symmetric black holes using what they called the 
Einstein-Euler-Heisenberg theory. \cite{2004TMP...140.1001D, 
2005PhRvD..71f3002D} looked at electromagnetic wave propagation in a strong 
gravitational field described by the Schwarzschild metric and including QED 
effects. In another series of papers, \cite{2003A&A...399L..39D, 
2014PhRvD..90b3011D} examined the propagation of light rays, bending and time 
delay due to vacuum birefringence.

From an observational point of view, X-ray polarization is the key 
observable to study QED effects in general relativity in the context of neutron 
star magnetospheres and atmospheres. Such signatures have indeed already been 
investigated by \cite{2002PhRvD..66b3002H}. They showed that the X-ray 
polarization degree is largely enhanced compared to classical field theory. 
\cite{2000MNRAS.311..555H} also demonstrated a possible phase lag between 
different wavelengths. \cite{2014MNRAS.438.1686T} showed that phase-resolved 
polarimetry gives insight into magnetar magnetospheres in the ultra strong 
field regime already with modest X-ray telescopes. At optical and infrared 
wavelengths, plasma and vacuum polarization effects compete, leading to possible 
constraints on the total charge density within the magnetosphere as explained 
by \cite{2006MNRAS.368.1377S}. Moreover because vacuum birefringence in strong 
magnetic field leads to variable refractive indexes, \cite{1999MNRAS.306..333S} 
predicted strong QED lensing analogue to the gravitational lensing although in 
QED lensing the strength of the lens depends on the polarization states of the 
photons \citep{2005PhRvL..94p1101D, 2012JCAP...10..056K}. Such effects are 
supposed to help diagnosing the magnetic field structure around neutron stars.

To study the magnetosphere of neutron stars, it is often useful to start with a 
simple approximation called force-free electrodynamics (FFE). It represents a 
zero order approximation to compute the plasma response and current density 
knowing the electromagnetic field topology. FFE has been used in the last decade 
by many authors to investigate pulsar and magnetar magnetospheres 
\citep{2006MNRAS.368.1055T, 2006ApJ...648L..51S, 2006MNRAS.368L..30M, 
2012MNRAS.424..605P, 2012MNRAS.423.1416P, 2015MNRAS.447.3170P}. 
\cite{2015arXiv150303867F} showed that quantum force-free electrodynamics can be 
recast into the 3+1 language as a set of time evolution equations for the 
electric and magnetic field. The system looks very similar to 
special-relativistic FFE except for some corrections in the current 
density.

In the birefringent medium induced by vacuum polarization and 
in the presence of plasma, vacuum resonances occurs, which translates into a 
mode conversion from low to high opacity, according to 
\cite{2003ApJ...588..962L}. But this resonance depends on the energy~$E$ of the 
photon, so it is important to include possible gravitational redshift effects 
into that picture, especially when the radiation is emitted from the 
neutron star surface. This vacuum resonance imprints a special signature in the 
X-ray polarization properties from its surface emission 
\citep{2003PhRvL..91g1101L}. Because the vacuum resonance condition includes~$E$ 
it has to be corrected for the strong gravitational field. A 3+1~formalism as 
the 
one we developed here might help to better understand such resonance in 
neutron star magnetospheres. For instance, our framework could be used to 
extend the Monte Carlo simulations for Compton scattering performed by 
\cite{1997MNRAS.288..596B} for soft gamma repeaters in flat space-time or by 
\cite{2011ApJ...730..131F} in a Schwarzschild background metric accounting for 
light ray deflection as proposed by \cite{2003MNRAS.342..134H}. It could also 
serve to compute the radiation spectra 
from vacuum polarization and proton cyclotron resonances as done by 
\cite{2003ApJ...583..402O} or atmospheres of neutron stars and their resulting 
spectra as explained by \cite{2003MNRAS.338..233H}. As a general conclusion, 
our approach combining gravity and QED in a 3+1 split of space-time will be 
extremely useful to straightforwardly extend the investigations cited before 
into the realm of general relativity. This will enable to make accurate 
predictions about the observational signature of neutron star atmospheres and 
magnetospheres.

In this paper, we include the most general background metric into the QED 
description. The 3+1~formalism used to 
split the covariant Maxwell equations into space and time is summarized in 
section~\ref{sec:Relativite}. Next the Lagrangian formalism for the 
electromagnetic field is defined in curved space-time in 
section~\ref{sec:Lagrangien}. In section~\ref{sec:NLE} we derive the non-linear 
electromagnetic field equations in curved space-time. An application to the 
spin-down luminosity in strongly 
magnetized neutron stars is discussed in section~\ref{sec:Application}. 
Concluding remarks are given in 
section~\ref{sec:Conclusion}.

\section{The 3+1 formalism in general relativity}
\label{sec:Relativite}

In order to apply traditional finite volume schemes and to get more physical 
insight into electromagnetism in general relativity, we remind for completeness 
the 3+1~language often used to transform covariant equations into time-dependent 
hyperbolic systems of spatial vectors in three dimensions. To this aim, we split 
the four dimensional space-time into a 3+1~foliation such that the metric can be 
expressed as
\begin{equation}
  \label{eq:metrique}
  ds^2 = \alpha^2 \, c^2 \, dt^2 - \gamma_{ab} \, ( dx^a + \beta^a \, c\,dt ) \, (dx^b + \beta^b \, c\,dt )
\end{equation}
where $x^i = (c\,t,x^a)$, $t$ is the time coordinate or universal time and $x^a$ 
some associated space coordinates. We use the Landau-Lifschitz convention for 
the metric signature given by $(+,-,-,-)$ \citep{LandauLifchitzTome2}. $\alpha$ 
is the lapse function, $\beta^a$ the shift vector and $\gamma_{ab}$ the spatial 
metric of absolute space. By convention, latine letters from $a$ to $h$ are used 
for the components of vectors in absolute space (in the range~$\{1,2,3\}$) 
whereas latine letters starting from $i$ are used for four dimensional vectors 
and tensors (in the range~$\{0,1,2,3\}$). Our derivation of the 3+1 equations 
follows the method outlined by \cite{2004MNRAS.350..427K, 2011MNRAS.418L..94K} 
and extensively used by \cite{2013MNRAS.433..986P, 2014MNRAS.439.1071P, 
2015MNRAS.447.3170P}.

Let $F^{ik}$ and ${^*F}^{ik}$ be the electromagnetic tensor and its dual 
respectively. It is useful to introduce the following spatial vectors $(\mathbf 
B, \mathbf E, \mathbf D, \mathbf H)$ such that
\begin{subequations}
\label{eq:BDEH}
\begin{align}
 B^a & = \alpha \, {^*F}^{a0} \\
 E_a & = \frac{\alpha}{2} \, e_{abc} \, c \, {^*F}^{bc} \\
 D^a & = \varepsilon_0 \, c \, \alpha \, F^{a0} \\
 H_a & = - \frac{\alpha}{2\,\mu_0} \, e_{abc} \, F^{bc}
\end{align}
\end{subequations}
where $\varepsilon_0$ is the vacuum permittivity and $\mu_0$ the vacuum 
permeability, $e_{abc} = \sqrt{\gamma} \, \varepsilon_{abc}$ the fully 
antisymmetric spatial tensor and $\varepsilon_{abc}$ the three dimensional 
Levi-Civita symbol. The contravariant analogue is $e^{abc} = 
\varepsilon^{abc}/\sqrt{\gamma}$. $\gamma$ is the determinant of the spatial 
metric $\gamma_{ab}$. The three dimensional vector fields are not independent, 
they are related by two important constitutive relations, namely 
\begin{subequations}
\label{eq:Constitutive}
\begin{align}
\label{eq:ConstitutiveE}
  \varepsilon_0 \, \mathbf E & = \alpha \, \mathbf D + \varepsilon_0\,c\,\mathbf\beta \times \mathbf B \\
\label{eq:ConstitutiveH}
  \mu_0 \, \mathbf H & = \alpha \, \mathbf B - \frac{\mathbf\beta \times \mathbf 
D}{\varepsilon_0\,c} \;.
\end{align}
\end{subequations}
The curvature of absolute space is taken into account by the lapse function factor~$\alpha$ in the first term on the right-hand side and the frame dragging effect is included in the second term, the cross-product between the shift vector~$\mathbf\beta$ and the fields. 

To complete the description of the electromagnetic field in general relativity 
including QED corrections, we need to derive the field equations. This 
is done starting from a Lagrangian for the electromagnetic field, as presented 
in the next section.

\section{Lagrangian of the electromagnetic field}
\label{sec:Lagrangien}

In classical field theory, the Lagrangian~$\mathcal{L}$ of the electromagnetic field is given according to the two field invariants expressed in covariant form as $\mathcal{I}_1 = F_{ik} \, F^{ik}$ and $\mathcal{I}_2 = F_{ik}\,{^*F}^{ik}$. To the lowest order in these invariants, the classical Lagrangian \citep{2001elcl.book.....J} is given by
\begin{equation}
 \mathcal{L}_0 = - \frac{1}{4\,\mu_0} \, F_{ik} \, F^{ik} - I^i \, A_i
\end{equation}
where $I^i$ is the four current and $A_i$ the four potential of the electromagnetic field. In general relativity, the evolution equations for the electromagnetic field follow then from the variational principle expressed by Euler-Lagrange equations \citep{uzan2014} such that
\begin{equation}
\label{eq:EulerLagrange}
 \frac{\partial \mathcal{L}}{\partial A_i} - \frac{1}{\sqrt{-g}} \, \partial_k \sqrt{-g} \, \frac{\partial \mathcal{L}}{\partial \partial_k A_i} = 0
\end{equation}
where~$g = \alpha \, \sqrt{\gamma}$ is the determinant of the space-time 
metric in equation~(\ref{eq:metrique}). Any non-linear covariant theory of 
electrodynamics starts from a Lagrangian including higher orders of the 
invariants~$\mathcal{I}_1$ and $\mathcal{I}_2$, like for instance the Lagrangian 
proposed by \cite{1936ZPhy...98..714H} or \cite{1934RSPSA.144..425B}. To remain 
as general as possible in our discussion, we use a parametrized post-Maxwellian 
description, similar to the parametrized post-Newtonian case used for the 
gravitational field. The lowest order correction includes two parameters 
$(\eta_1,\eta_2)$ such that the Lagrangian to this order becomes
\begin{equation}
\label{eq:Lagrangien}
 \mathcal{L} = \mathcal{L}_0 + \eta_1 \, \mathcal{I}_1^2 + \eta_2 \, 
\mathcal{I}_2^2 \;.
\end{equation}
In the Euler-Heisenberg prescription we have
\begin{subequations}
\begin{align}
 \eta_1 & = \frac{\alpha_{\rm sf}}{180\,\upi} \, \frac{1}{2\,\mu_0\,\BQ^2} \\
 \eta_2 & = \frac{7}{4} \, \eta_1
\end{align}
\end{subequations}
whereas for the Born-Infeld Lagrangian in the weak field limit we have
\begin{subequations}
\begin{align}
 \eta_1 & = \frac{1}{32\,\mu_0\,b^2} \\
 \eta_2 & = \eta_1
\end{align}
\end{subequations}
with $\alpha_{\rm sf}$ the fine structure constant and
$b=9.18\times10^{11}$~T the empirical maximal absolute field strength in 
Born-Infeld theory.

Comparing the relative strength of the classical and QED Lagrangian, we deduce that quantum corrections to~$\mathcal{L}$ are of the order
\begin{equation}
 \frac{\mathcal{L} - \mathcal{L}_0}{\mathcal{L}_0} \approx 
4\,\mu_0\,\eta_1F_{ik} \, F^{ik} \approx \frac{\alpha_{\rm sf}}{90\,\upi} \, 
\frac{B^2}{\BQ^2} \approx 2.6\times10^{-5} \, \frac{B^2}{\BQ^2}
\end{equation}
and therefore remain small even in the case of magnetic fields close to the 
critical field~$\BQ$. QED can always be treated as a small non-linear 
perturbation of Maxwell equation. For magnetic fields well above $\BQ$ the 
aforementioned perturbative Lagrangians do not hold any longer but it can be 
shown that the relative strength between classical and QED Lagrangian remains 
much less than unity \citep{1997PhRvD..55.2449H, 2006hep.th....9081R} unless $B$ 
reaches unrealistic value of the order $\BQ \, e^{3\,\upi/\alpha_{\rm sf}} 
\approx 10^{570}$~T because of the logarithmic dependence in $\ln (B/\BQ)$ of 
the Lagrangian in this ultra high magnetic field limit 
\citep{LandauLifchitzTome4}.

\section{Non linear electrodynamics in general relativity}
\label{sec:NLE}

From the above first order corrections to the Lagrangian~$\mathcal{L}$, 
eq.~(\ref{eq:Lagrangien}), it is straightforward to get the time evolution of 
the electromagnetic field, that is the non-linear Maxwell equations in general 
relativity.

\subsection{Maxwell equations}

The equation of motion for the fields are given by the expression~(\ref{eq:EulerLagrange}). To write them down in the 3+1 language we note that
\begin{equation}
 \frac{\partial \mathcal{L}}{\partial \partial_l A_m} = - \frac{F^{lm}}{\mu_0} + 8 \, \eta_1 \, F^{lm} \, \mathcal{I}_1 + 8 \, \eta_2 \, {^*F}^{lm} \, \mathcal{I}_2
\end{equation}
or briefly with the spatial vectors $\bmath{D}$ and $\bmath{B}$
\begin{subequations}
 \begin{align}
 \frac{\partial \mathcal{L}}{\partial \partial_l A_m} & = - \xi_1 \, \frac{F^{lm}}{\mu_0} - \xi_2 \, {^*F}^{lm} \\
 \xi_1 & = 1 - 16\,\mu_0\,\eta_1 \, \left( B^2 - \frac{\mu_0}{\varepsilon_0} \, D^2 \right) \\
 \xi_2 & = 32 \, \eta_2 \, \frac{\bmath{D} \cdot \bmath{B}}{\varepsilon_0\,c} 
\;.
 \end{align}
\end{subequations}
Following the standard 3+1 decomposition used in general relativity we introduce 
two more auxiliary vector fields denoted by $\bmath{F}$ and $\bmath{G}$ and 
defined by 
\begin{subequations}
\label{eq:FGDBEH}
 \begin{align}
\label{eq:FDB}
 \bmath{F} & = \xi_1 \, \bmath{D} + \frac{\xi_2}{c} \, \bmath{B} \\
\label{eq:GEH}
 \bmath{G} & = \xi_1 \, \bmath{H} - \frac{\xi_2}{c} \, \bmath{E} \;.
 \end{align}
\end{subequations}
Straightforward computation shows that
\begin{subequations}
 \begin{align}
  \frac{\partial \mathcal{L}}{\partial A_i} & = - I^i \\
  \frac{1}{\sqrt{-g}} \, \partial_k \sqrt{-g} \, \frac{\partial \mathcal{L}}{\partial \partial_k A_0} & = \nabla_k \left( \xi_1 \, \frac{F^{0k}}{\mu_0} + \xi_2 \, {^*F}^{0k} \right) \\
   & = - \frac{c}{\alpha} \, \divg\bmath{F} \\
  \frac{1}{\sqrt{-g}} \, \partial_k \sqrt{-g} \, \frac{\partial \mathcal{L}}{\partial \partial_k A_a} & = \nabla_k \left( \xi_1 \, \frac{F^{ak}}{\mu_0} + \xi_2 \, {^*F}^{ak} \right) \\
  & = \frac{1}{\alpha\,\sqrt{\gamma}} \, \partial_t \left( \sqrt{\gamma} \, 
\bmath{F} \right) - \frac{1}{\alpha} \, \rot \bmath{G} \;.
 \end{align}
\end{subequations}
The inhomogeneous Maxwell equations then follow immediately as
\begin{subequations}
\label{eq:GRQEDMaxwell}
 \begin{align}
\label{eq:MaxwellInhomo1}
 \divg \bmath{F} & = \rho \\
\label{eq:MaxwellInhomo2}
 \rot \bmath{G} & = \bmath{J} + \frac{1}{\sqrt{\gamma}} \, \partial_t 
(\sqrt{\gamma} \, \bmath{F}) \;.
 \end{align}
Note that $\rho=\alpha\,I^0/c$ has to be interpreted as the external charge 
density and $J^a=\alpha\,I^a$ as the external current density. Vacuum 
polarization is treated implicitly through the definition of the vectors 
$\bmath{F}$ and $\bmath{G}$ as given by equations~(\ref{eq:FGDBEH}).
The homogeneous Maxwell equations are not modified, they read as in classical general relativity
\begin{align}
  \label{eq:Maxwell_Faraday}
  \rot \bmath{E} & = - \frac{1}{\sqrt{\gamma}} \, \partial_t (\sqrt{\gamma} \, \bmath{B} ) \\
  \label{eq:Div_B_0}
  \divg \bmath{B} & = 0 \;.
\end{align}
\end{subequations}
It is seen from eq.~(\ref{eq:GRQEDMaxwell}) that the primary fields to be 
evolved are $\bmath{B}$ and $\bmath{F}$. The other auxiliary fields are then 
deduced from the constitutive relations. Indeed, $\bmath{B}$ and  $\bmath{F}$ 
being known, we can retrieve $\bmath{D}$ from eq.~(\ref{eq:FDB}). Next from 
$\bmath{D}$ and $\bmath{B}$ we can get $\bmath{E}$ and $\bmath{H}$ through 
eq.~(\ref{eq:Constitutive}). Finally $\bmath{G}$ is obtained from 
eq.~(\ref{eq:GEH}) knowing $\bmath{E}$ and $\bmath{H}$ from the previous 
calculation. This completes one full time step to advance the primary fields 
$\bmath{B}$ and $\bmath{F}$.

The source terms $\rho$ and $\bmath{J}$ are left free so far. In quantum vacuum as well as in classical vacuum, they vanish $\rho=0$ and $\bmath{J}=\bmath{0}$. In the most general case, source terms have to be specified by some other equations, like the conservation of energy-momentum of the plasma. Nevertheless, in some restricted problems, it is possible to compute the current density from only the knowledge of the electromagnetic field. Such an example, called force-free electrodynamics is presented in the next section.

\subsection{Force-free quantum electrodynamics (FFQED)}

The source terms have not yet been specified. Having in mind to apply the above 
equations to pulsar and magnetar magnetospheres, we give expressions for the 
current density in the limit of a force-free plasma, neglecting inertia and 
pressure. The force-free condition in covariant form and including QED 
corrections reads
\begin{equation}
 F_{ik} \, I^k = 0
\end{equation}
and in the 3+1~formalism it becomes
\begin{subequations}
\begin{align}
  \mathbf J \cdot \mathbf E & = 0 \\
  \rho \, \mathbf E + \mathbf J \times \mathbf B & = \mathbf 0
\end{align}
\end{subequations}
which implies $\mathbf E \cdot \mathbf B = 0 $ and therefore also $\mathbf D 
\cdot \mathbf B = 0 $. Thus the parameter $\xi_2$ must vanish and therefore 
it follows that $\bmath{F} \cdot \bmath{B}=0$. Applying the usual technique from 
special relativistic electrodynamics we get the current density including  
general relativity and QED such that
\begin{equation}
 \bmath{J} = \rho \, \frac{\bmath{E} \wedge \bmath{B}}{B^2} + \frac{\bmath{B} \cdot \rot \bmath{G} - \bmath{F} \cdot \rot \bmath{E}}{B^2} \, \bmath{B} .
\end{equation}
Because $c\,B^a={^*F}^{ak}\,n_k$ and $D^a/\varepsilon_0 = F^{ak}\,n_k$, $\mathbf 
B$ and $\mathbf D/\varepsilon_0$ can be interpreted as the magnetic and electric 
field respectively as measured by the fiducial observer whose four velocity 
is $n_k=(\alpha\,c,\bmath 0)$. Moreover
\begin{equation}
  \label{eq:DensiteFIDO}
  I^k \, n_k = \rho \, c^2  
\end{equation}
thus $\rho$ is the electric charge density as measured by this same observer. Its electric current density~$\mathbf j$ is given by
\begin{equation}
  \label{eq:CourantFIDO}
  \alpha \, \mathbf j = \mathbf J + \rho \, c \, \mathbf \beta \;.
\end{equation}
Maxwell equations~(\ref{eq:GRQEDMaxwell}), the constitutive relations 
(\ref{eq:Constitutive}),(\ref{eq:FGDBEH}) and the prescription for the source 
terms set the background system to be solved for any prescribed metric in the 
low electromagnetic field limit according to the classical Lagrangian 
correction. As a check of these equations, let us compute their simplified 
version in the classical general-relativistic as well as in the 
special-relativistic QED limits.

\subsection{Classical general-relativistic limit}

If QED corrections are neglected, we have to set $\xi_1=1$ and $\xi_2=0$. In 
this limit the new auxiliary fields $\bmath{F}$ and $\bmath{G}$ reduce 
respectively to the classical fields $\bmath{D}$ and $\bmath{H}$. The 
inhomogeneous Maxwell equations become
\begin{subequations}
\label{eq:GRMaxwellInhomo}
 \begin{align}
 \divg \bmath{D} & = \rho \\
 \rot \bmath{H} & = \bmath{J} + \frac{1}{\sqrt{\gamma}} \, \partial_t (\sqrt{\gamma} \, \bmath{D})
 \end{align}
\end{subequations}
which are the expressions found by \cite{2004MNRAS.350..427K, 
2011MNRAS.418L..94K} and by \cite{2013MNRAS.433..986P}. The current 
density then simplifies to 
\begin{equation}
 \mathbf J = \rho \, \frac{\mathbf E \times \mathbf B}{B^2} + \frac{\mathbf B \cdot \rot \mathbf H - \mathbf D \cdot \rot \mathbf E}{B^2} \, \mathbf B
\end{equation}
as expected from the above cited works.

\subsection{Quantum special-relativistic limit}

In the other limit, when the gravitational field becomes negligible, the lapse 
function is equal to one, $\alpha=1$, and the shift vector vanishes, 
$\bmath{\beta}=\bmath{0}$. The constitutive relations  (\ref{eq:Constitutive}) 
simplify 
to $\varepsilon_0\,\bmath{E}=\bmath{D}$ and $\mu_0\,\bmath{H}=\bmath{B}$. 
Maxwell equations then resemble those in a medium given by
\begin{subequations}
\label{eq:QSRMaxwell}
\begin{align}
  \rot \bmath{E} & = - \frac{\partial\bmath{B}}{\partial t} \\
  \label{eq:Maxwell_Ampere}
  \rot \bmath{G} & = \bmath{J} + \frac{\partial\bmath{F}}{\partial t}
 \end{align}
with the initial conditions on the divergence of the $\bmath{F}$ and 
$\bmath{B}$ fields such as
\begin{align}
  \divg \bmath{B} & = 0 \\
  \label{eq:Div_E_Rho}
  \divg \bmath{F} & = \rho \;.
\end{align}
\end{subequations}
To retrieve the more familiar notation, we should substitute $\bmath{F}$ by 
$\bmath{D}$ and $\bmath{G}$ by $\bmath{H}$. Electrodynamics in the presence of 
strong electromagnetic fields can be described by non linear Maxwell equations 
derived from an effective Lagrangian computed by Euler and Heisenberg. Quantum 
electrodynamics described vacuum as a polarized and magnetized media without 
external current density,~$\bmath{J}=\bmath{0}$, and without any charge 
density,~$\rho=0$. The quantum vacuum is depicted by a magnetization~$\bmath M$ 
and a polarisation~$\bmath P$ such that 
\begin{subequations}
\label{eq:Constitutives}
 \begin{align}
 \bmath{F} & = \varepsilon_0 \, \bmath{E} + \bmath{P} \\
 \bmath{G} & = \bmath{B}/\mu_0 - \bmath{M} \;.
\end{align}
\end{subequations}
To the first order in the fine structure constant, the Euler and Heisenberg Lagrangian shows that
\begin{subequations}
 \begin{align}
  \bmath P & = \kappa \, ( 2 \, ( E^2 - c^2\,B^2 ) \, \bmath E + 7 \, c^2 \, (\bmath{E} \cdot \bmath{B}) \, \bmath B ) \\
  \bmath M & = - \kappa \,  ( 2 \, c^2\,( E^2 - c^2\,B^2 ) \, \bmath B - 7 \, c^2\,(\bmath{E} \cdot \bmath{B}) \, \bmath E )
 \end{align}
\end{subequations}
with
\begin{equation}
 \kappa = \frac{\alpha_{\rm sf}}{45\,\upi\,\mu_0\,c^4\,\BQ^2} \;.
\end{equation}
Consequently, our constitutive relations include both limits, the classical 
general-relativistic field and QED in Newtonian gravitational field. We refer 
to it as general-relativistic force-free quantum electrodynamics (GRFFQED).

\section{Spin-down luminosity of pulsars and magnetars}
\label{sec:Application}

What kind of change in the spin-down luminosity can we expect from the vacuum 
polarization? We know already from previous works by \cite{2004MNRAS.352.1161R} 
and \cite{2013MNRAS.433..986P} that general relativity leads to an increase by a 
factor 2 to 6 of the inferred spin-down losses compared to flat space-time. 
Nevertheless Maxwell equations remain linear in general relativity. The enhanced 
luminosity can be interpreted as a combination of magnetic field amplification 
and gravitational redshift of the angular frequency of the neutron star due to 
the curvature of space-time. Moreover, from the lowest order corrections 
brought 
by QED, Maxwell field equations become non-linear. These non-linearities will 
perturb the topology of the dipole but they will also generate higher order 
multipoles like for instance an hexapole due to terms containing products 
of $(E^2,B^2)$ with $(\bmath{B,E})$. For the corrections to the dipole we have 
\begin{equation}
 \frac{\delta L_{\rm dip}}{L_{\rm dip}} = 2 \, \frac{\delta B}{B} \approx \frac{4\,\alpha_{\rm sf}}{45\,\upi} \, \frac{B^2}{\BQ^2} \approx 3 \times 10^{-4} \, \frac{B^2}{\BQ^2}
\end{equation}
which remains negligible. These conclusions have already been found by 
\cite{1997JPhA...30.6475H}. For the hexapole, using the point 
multipole formula for the most luminous mode i.e. $m=2$ given by 
\cite{2015MNRAS.450..714P} we find
\begin{equation}
 \frac{L_{\rm hex}}{L_{\rm dip}} = \frac{243}{25} \, \left( \frac{R}{\rlight} \right)^4 \, \frac{\delta B^2}{B^2} \ll \frac{\delta B^2}{B^2} \approx 10^{-7} \, \frac{B^4}{\BQ^4}
\end{equation}
so even less significant than the dipole corrections. We conclude that whatever 
the period and the magnetic field strength of pulsars and magnetars, QED 
corrections to the spin-down luminosities of any multipolar component remain 
meaningless with respect to the corrections brought by general relativity 
alone.

\section{Conclusion}
\label{sec:Conclusion}

In this paper we showed how to include vacuum polarization into the description 
of a force-free magnetosphere in general relativity according to the weak 
electromagnetic field limit i.e. in the first order perturbation theory of the 
Lagrangian for the electromagnetic field. By introducing two new auxiliary 
fields, it is possible to cast the full set of Maxwell equations into a 
classical three dimensional picture treating curved space-time and vacuum 
polarization as two medium with specified constitutive relations for both parts. 
Our new formalism should help to quantify the merit of both contributions, 
general relativity and quantum electrodynamics, to the dynamics of neutron star 
magnetospheres, especially for magnetars and pulsars with high-B fields. As 
shown in the previous section, on a global length scale related to the 
size of the magnetosphere and to its rotational braking, QED is irrelevant 
to account for variation in the spin-down luminosity. Although QED corrections 
to the electrodynamics of neutron star magnetosphere remain weak, its 
implications for interpretation of observations like propagation and 
polarization of electromagnetic waves and pair creation are likely to be 
important.

\section*{Acknowledgements}

I am grateful to the referee for helpful comments and suggestions. This 
work has been supported by the French National Research Agency (ANR) through 
the grant No. ANR-13-JS05-0003-01 (project EMPERE).


\begin{thebibliography}{56}
\expandafter\ifx\csname natexlab\endcsname\relax\def\natexlab#1{#1}\fi

\bibitem[{{Battesti} \& {Rizzo}(2013)}]{2013RPPh...76a6401B}
{Battesti} R., {Rizzo} C., 2013, Reports on Progress in Physics, 76, 016401

\bibitem[{{Born} \& {Infeld}(1934)}]{1934RSPSA.144..425B}
{Born} M., {Infeld} L., 1934, Royal Society of London Proceedings Series A,
  144, 425

\bibitem[{{Bulik} \& {Miller}(1997)}]{1997MNRAS.288..596B}
{Bulik} T., {Miller} M.~C., 1997, \mnras, 288, 596

\bibitem[{{Denisov} {et~al}\mbox{.}(2004){Denisov}, {Denisova}, \&
  {Svertilov}}]{2004TMP...140.1001D}
{Denisov} V.~I., {Denisova} I.~P., {Svertilov} S.~I., 2004, Theoretical and
  Mathematical Physics, 140, 1001

\bibitem[{{Denisov} {et~al}\mbox{.}(2014){Denisov}, {Sokolov}, \&
  {Vasili'ev}}]{2014PhRvD..90b3011D}
{Denisov} V.~I., {Sokolov} V.~A., {Vasili'ev} M.~I., 2014, \prd, 90, 023011

\bibitem[{{Denisov} \& {Svertilov}(2003)}]{2003A&A...399L..39D}
{Denisov} V.~I., {Svertilov} S.~I., 2003, \aap, 399, L39

\bibitem[{{Denisov} \& {Svertilov}(2005)}]{2005PhRvD..71f3002D}
{Denisov} V.~I., {Svertilov} S.~I., 2005, \prd, 71, 063002

\bibitem[{{Dupays} {et~al}\mbox{.}(2005){Dupays}, {Robilliard}, {Rizzo}, \&
  {Bignami}}]{2005PhRvL..94p1101D}
{Dupays} A., {Robilliard} C., {Rizzo} C., {Bignami} G.~F., 2005, Physical
  Review Letters, 94, 161101

\bibitem[{{Fern{\'a}ndez} \& {Davis}(2011)}]{2011ApJ...730..131F}
{Fern{\'a}ndez} R., {Davis} S.~W., 2011, \apj, 730, 131

\bibitem[{{Freytsis} \& {Gralla}(2015)}]{2015arXiv150303867F}
{Freytsis} M., {Gralla} S.~E., 2015, ArXiv e-prints

\bibitem[{{Haas}(2005)}]{2005PhPl...12f2117H}
{Haas} F., 2005, Physics of Plasmas, 12, 062117

\bibitem[{{Harding} \& {Lai}(2006)}]{2006RPPh...69.2631H}
{Harding} A.~K., {Lai} D., 2006, Reports on Progress in Physics, 69, 2631

\bibitem[{{Heisenberg} \& {Euler}(1936)}]{1936ZPhy...98..714H}
{Heisenberg} W., {Euler} H., 1936, Zeitschrift fur Physik, 98, 714

\bibitem[{{Heyl}(2001)}]{2001PhRvD..63f4028H}
{Heyl} J.~S., 2001, \prd, 63, 064028

\bibitem[{{Heyl}(2007)}]{2007Ap&SS.308..101H}
{Heyl} J.~S., 2007, \apss, 308, 101

\bibitem[{{Heyl} \& {Hernquist}(1997{\natexlab{a}})}]{1997PhRvD..55.2449H}
{Heyl} J.~S., {Hernquist} L., 1997{\natexlab{a}}, \prd, 55, 2449

\bibitem[{{Heyl} \& {Hernquist}(1997{\natexlab{b}})}]{1997JPhA...30.6475H}
{Heyl} J.~S., {Hernquist} L., 1997{\natexlab{b}}, Journal of Physics A
  Mathematical General, 30, 6475

\bibitem[{{Heyl} \& {Hernquist}(1998)}]{1998PhRvD..58d3005H}
{Heyl} J.~S., {Hernquist} L., 1998, \prd, 58, 043005

\bibitem[{{Heyl} \& {Hernquist}(1999)}]{1999PhRvD..59d5005H}
{Heyl} J.~S., {Hernquist} L., 1999, \prd, 59, 045005

\bibitem[{{Heyl} \& {Hernquist}(2005)}]{2005MNRAS.362..777H}
{Heyl} J.~S., {Hernquist} L., 2005, \mnras, 362, 777

\bibitem[{{Heyl} \& {Shaviv}(2000)}]{2000MNRAS.311..555H}
{Heyl} J.~S., {Shaviv} N.~J., 2000, \mnras, 311, 555

\bibitem[{{Heyl} \& {Shaviv}(2002)}]{2002PhRvD..66b3002H}
{Heyl} J.~S., {Shaviv} N.~J., 2002, \prd, 66, 023002

\bibitem[{{Heyl} {et~al}\mbox{.}(2003){Heyl}, {Shaviv}, \&
  {Lloyd}}]{2003MNRAS.342..134H}
{Heyl} J.~S., {Shaviv} N.~J., {Lloyd} D., 2003, \mnras, 342, 134

\bibitem[{{Ho} \& {Lai}(2003)}]{2003MNRAS.338..233H}
{Ho} W.~C.~G., {Lai} D., 2003, \mnras, 338, 233

\bibitem[{{Jackson}(2001)}]{2001elcl.book.....J}
{Jackson} J.~D., 2001, {\'Electrodynamique classique}. Dunod, 2001

\bibitem[{{Kim}(2012)}]{2012JCAP...10..056K}
{Kim} J.~Y., 2012, \jcap, 10, 56

\bibitem[{{King} \& {Di Piazza}(2014)}]{2014EPJST.223.1063K}
{King} B., {Di Piazza} A., 2014, European Physical Journal Special Topics, 223,
  1063

\bibitem[{{Komissarov}(2004)}]{2004MNRAS.350..427K}
{Komissarov} S.~S., 2004, \mnras, 350, 427

\bibitem[{{Komissarov}(2011)}]{2011MNRAS.418L..94K}
{Komissarov} S.~S., 2011, \mnras, 418, L94

\bibitem[{{Lai} \& {Ho}(2003{\natexlab{a}})}]{2003PhRvL..91g1101L}
{Lai} D., {Ho} W.~C., 2003{\natexlab{a}}, Physical Review Letters, 91, 071101

\bibitem[{{Lai} \& {Ho}(2003{\natexlab{b}})}]{2003ApJ...588..962L}
{Lai} D., {Ho} W.~C.~G., 2003{\natexlab{b}}, \apj, 588, 962

\bibitem[{{Landau} \& {Lifchitz}(1989{\natexlab{a}})}]{LandauLifchitzTome4}
{Landau} L., {Lifchitz} E., 1989{\natexlab{a}}, \emph{\'Electrodynamique
  quantique}. Editions MIR Moscou

\bibitem[{{Landau} \& {Lifchitz}(1989{\natexlab{b}})}]{LandauLifchitzTome2}
{Landau} L., {Lifchitz} E., 1989{\natexlab{b}}, \emph{Th\'eorie des champs}.
  Editions MIR Moscou

\bibitem[{{Manfredi} \& {Haas}(2001)}]{2001PhRvB..64g5316M}
{Manfredi} G., {Haas} F., 2001, \prb, 64, 075316

\bibitem[{{Mazur} \& {Heyl}(2011)}]{2011MNRAS.412.1381M}
{Mazur} D., {Heyl} J.~S., 2011, \mnras, 412, 1381

\bibitem[{{McKinney}(2006)}]{2006MNRAS.368L..30M}
{McKinney} J.~C., 2006, \mnras, 368, L30

\bibitem[{{Olausen} \& {Kaspi}(2014)}]{2014ApJS..212....6O}
{Olausen} S.~A., {Kaspi} V.~M., 2014, \apjs, 212, 6

\bibitem[{{{\"O}zel}(2003)}]{2003ApJ...583..402O}
{{\"O}zel} F., 2003, \apj, 583, 402

\bibitem[{{Parfrey} {et~al}\mbox{.}(2012){Parfrey}, {Beloborodov}, \&
  {Hui}}]{2012MNRAS.423.1416P}
{Parfrey} K., {Beloborodov} A.~M., {Hui} L., 2012, \mnras, 423, 1416

\bibitem[{{P{\'e}tri}(2012)}]{2012MNRAS.424..605P}
{P{\'e}tri} J., 2012, \mnras, 424, 605

\bibitem[{{P{\'e}tri}(2013)}]{2013MNRAS.433..986P}
{P{\'e}tri} J., 2013, \mnras, 433, 986

\bibitem[{{P{\'e}tri}(2014)}]{2014MNRAS.439.1071P}
{P{\'e}tri} J., 2014, \mnras, 439, 1071

\bibitem[{{P{\'e}tri}(2015{\natexlab{a}})}]{2015MNRAS.447.3170P}
{P{\'e}tri} J., 2015{\natexlab{a}}, \mnras, 447, 3170

\bibitem[{{P{\'e}tri}(2015{\natexlab{b}})}]{2015MNRAS.450..714P}
{P{\'e}tri} J., 2015{\natexlab{b}}, \mnras, 450, 714

\bibitem[{{Rezzolla} \& {Ahmedov}(2004)}]{2004MNRAS.352.1161R}
{Rezzolla} L., {Ahmedov} B.~J., 2004, \mnras, 352, 1161

\bibitem[{{Ruffini} {et~al}\mbox{.}(2010){Ruffini}, {Vereshchagin}, \&
  {Xue}}]{2010PhR...487....1R}
{Ruffini} R., {Vereshchagin} G., {Xue} S.-S., 2010, \physrep, 487, 1

\bibitem[{{Ruffini} {et~al}\mbox{.}(2013){Ruffini}, {Wu}, \&
  {Xue}}]{2013PhRvD..88h5004R}
{Ruffini} R., {Wu} Y.-B., {Xue} S.-S., 2013, \prd, 88, 085004

\bibitem[{{Ruffini} \& {Xue}(2006)}]{2006hep.th....9081R}
{Ruffini} R., {Xue} S.-S., 2006, J.Korean Phys.Soc., 49, S715

\bibitem[{{Shannon} \& {Heyl}(2006)}]{2006MNRAS.368.1377S}
{Shannon} R.~M., {Heyl} J.~S., 2006, \mnras, 368, 1377

\bibitem[{{Shaviv} {et~al}\mbox{.}(1999){Shaviv}, {Heyl}, \&
  {Lithwick}}]{1999MNRAS.306..333S}
{Shaviv} N.~J., {Heyl} J.~S., {Lithwick} Y., 1999, \mnras, 306, 333

\bibitem[{{Spitkovsky}(2006)}]{2006ApJ...648L..51S}
{Spitkovsky} A., 2006, \apjl, 648, L51

\bibitem[{{Taverna} {et~al}\mbox{.}(2014){Taverna}, {Muleri}, {Turolla},
  {Soffitta}, {Fabiani}, \& {Nobili}}]{2014MNRAS.438.1686T}
{Taverna} R., {Muleri} F., {Turolla} R., {Soffitta} P., {Fabiani} S., {Nobili}
  L., 2014, \mnras, 438, 1686

\bibitem[{{Thompson} \& {Blaes}(1998)}]{1998PhRvD..57.3219T}
{Thompson} C., {Blaes} O., 1998, \prd, 57, 3219

\bibitem[{{Timokhin}(2006)}]{2006MNRAS.368.1055T}
{Timokhin} A.~N., 2006, \mnras, 368, 1055

\bibitem[{{Uzan} \& {Deruelle}(2014)}]{uzan2014}
{Uzan} J., {Deruelle} N., 2014, Th\'eories de la Relativit\'e. Belin

\bibitem[{{Uzdensky} \& {Rightley}(2014)}]{2014RPPh...77c6902U}
{Uzdensky} D.~A., {Rightley} S., 2014, Reports on Progress in Physics, 77,
  036902

\end{thebibliography}

\label{lastpage}

\end{document}